\documentclass[useAMS,usenatbib]{mnras}
 \usepackage{graphicx}
 \usepackage{graphics}
 \usepackage{multicol}
 \usepackage{subfig}
 \usepackage{etex}
 \usepackage{setspace}
 \usepackage{float}
 \usepackage{pdflscape}
 \usepackage{natbib}

\def\aj{AJ}%
\def\araa{ARA\&A}%
\def\apj{ApJ}%
\def\apjl{ApJ}%
\def\apjs{ApJS}%
\def\ao{Appl.~Opt.}%
\def\apss{Ap\&SS}%
\def\aap{A\&A}%
\def\mnras{MNRAS}%
\def\pasj{PASJ}%
\def\nat{Nature}%
\title[AstroSat observations of Swift~J0243.6+6124]{AstroSat Observations of the first Galactic ULX Pulsar Swift~J0243.6+6124}
 \author[A. Beri et al.]
{Aru~Beri,$^{1,5}$
\thanks{a.beri@soton.ac.uk}
 Sachindra Naik,$^{2}$
 K.P Singh,$^{1}$
 Gaurava K. Jaisawal,$^{3}$
 Sudip~Bhattacharyya,$^{4}$
 \newauthor
 Philip Charles,$^{5}$
 Wynn C. G. Ho,$^{6,7}$
 Chandreyee~Maitra,$^8$
 Dipankar~Bhattacharya,$^9$
 \newauthor
 Gulab C Dewangan,$^9$
 Matthew Middleton,$^{5}$
 Diego~Altamirano,$^{5}$
 Poshak~Gandhi,$^{5}$
 \newauthor
 Harsha Raichur$^{10}$
 \\
 $^1$, DST-INSPIRE Faculty, Indian Institute of Science Education and Research (IISER) Mohali, 
Punjab 140306, India  \\
$^2$,~Astronomy and Astrophysics Division, Physical Research Laboratory, Navrangapura, Ahmedabad-380009, Gujarat, India \\
$^3$, National Space Institute, Technical University of Denmark, Elektrovej 327-328, DK-2800 Lyngby, Denmark \\ 
$^4$, Department of Astronomy and Astrophysics, Tata Institute of Fundamental Research, Homi Bhabha Road, Mumbai 400005, India.\\
$^{5}$, Physics \& Astronomy, University of Southampton, Southampton, Hampshire SO17 1BJ, UK \\
$^6$,~Department of Physics and Astronomy, Haverford College, 370 Lancaster Avenue, Haverford, PA 19041, US \\
$^7$,~Mathematical Sciences and STAG Research Centre,
 University of Southampton, Southampton SO17 1BJ, UK \\
$^8$, Max-Planck-Institut f\"{u}r extraterrestrische Physik, Giessenbachstra{\ss}e 1, 85748 Garching, Germany \\
$^9$, Inter-University Center for Astronomy and Astrophysics, Ganeshkhind, Pune 411007, India \\
$^{10}$,~Nordita, KTH Royal Institute of Technology and Stockholm University, Rosalagstullsbacken, 23, SE-10691 Stockholm \\}

\begin{document}
\pagerange{\pageref{firstpage}--\pageref{lastpage}} 
\maketitle
\label{firstpage}
 
\begin{abstract}
	Swift~J0243.6+6124, the first Galactic ultra-luminous X-ray pulsar, was observed during its 2017-2018 outburst with \emph{AstroSat}
at both sub- and super-Eddington levels of accretion~with X-ray luminosities of $L_{X}~{\sim}~7{\times}10^{37}$ and $6{\times}10^{38}$~$erg~s^{-1}$, respectively.~Our broadband timing and spectral observations show that X-ray pulsations at ${\sim}~9.85~\rm{s}$ have been detected up to 150~keV when the source was accreting at the super-Eddington level.~The pulse profiles are a strong function of both energy and source luminosity,~showing a double-peaked profile with pulse fraction increasing from $\sim$~$10{\%}$ at $1.65~\rm{keV}$ to 40--80~$\%$ at $70~\rm{keV}$.~The continuum X-ray spectra are well-modeled with a high energy cut-off power law~($\Gamma$~${\sim}$~0.6-0.7) and one or two blackbody components with temperatures of $\sim$~0.35~$\rm{keV}$ and $1.2~\rm{keV}$, depending on the accretion level.~No iron line emission is observed at sub-Eddington level, while a broad emission feature at around 6.9~keV is observed at the super-Eddington level, along with a blackbody radius~($121-142~\rm{km}$) that indicates the presence of optically thick outflows.

\end{abstract}


\section{Introduction}

Ultra-luminous X-ray sources~(ULXs) are non-nuclear point-like objects
with apparent luminosities exceeding $10^{39} {\rm{erg~s^{-1}}}$.
A majority of the ULXs are found in external galaxies and
are often considered promising candidates to host heavier than stellar-mass black holes \citep[for a review see][]{Kaaret17}.
Coherent X-ray pulsations were discovered from
a ULX in M82, thanks to the fast timing capability of \emph{NuSTAR} \citep{Bachetti14}, 
making it the first
Ultra-luminous X-ray pulsar~(ULP).~Currently only a handful of ULPs are known:
M82~X--2 \citep{Bachetti14}, NGC~7793~P13 \citep{Furst16,Israel17a}, NGC~5907~ULX1 \citep{Israel17a},
 NGC~300~ULX1 \citep{Carpano18},  NGC~1313~X--2 \citep{Sathyaprakash2019}, ULX--7 in M51 \citep{Castillo2020}.  \\

A new transient X-ray source, Swift~J0243.6+6124~(hereafter, J0243) was detected 
in outburst by \emph{Swift}-\textsc{BAT} on October 3, 2017 \citep{Cenko17} and
X-ray pulsations at $\sim$~9.86~s were detected with \emph{Swift}-\textsc{XRT} in the 0.2-10~keV band \citep{Kennea17}.
Later, these pulsations were confirmed in the data from \emph{Fermi}~\textsc{GBM} \citep{Jenke17},
\emph{Swift}-\textsc{XRT} \citep{Beardmore17}, and \emph{NuSTAR} \citep{Bahramian17, Jaisawal18}. 
This outburst lasted for about five months, and several multi-wavelength
observations were performed from radio to hard X-rays.
The optical spectroscopic observations performed by \citet{Kouro17}
revealed that the optical counterpart in the system is a late Oe- or early
Be-type star. Later, \citet{Bikmaev17} confirmed the Be/X-ray binary (BeXRB)
nature of the source. The peak X-ray flux~($F_{peak}$) observed during the 2017 outburst of
J0243 is $\sim$ $7 \times 10^{-7}\rm{erg~cm^{-2}~s^{-1}}$ in the 3-80~keV energy band \citep{Doroshenko18}.
The source distance has been estimated independently using X-ray and optical
constraints. 
\emph{Gaia} gives the source distance~({\rm d})~$\sim$~$7_{-1.2}^{+1.5} \rm{kpc}$ \citep[][hereafter WH18]{WH18},
 from which the peak X-ray luminosity~($L_{X}$) during the giant outburst 
is found to be $\sim$~$5\times10^{39}\rm{erg~s^{-1}}$ in the 3--80~keV band \citep{Tsygankov18}.
This peak $L_{X}$ exceeds the Eddington limit for a neutron star~(NS)
by a factor of $\sim$ 40,~thus, making J0243 the first Galactic X-ray pulsar to belong to the
recently discovered family of ULPs.
\emph{NuSTAR} observed J0243 several times during its outburst.~The results from the broadband 
spectroscopy in the 3-79~keV band revealed the presence of
a high-temperature black-body~($\rm {kT}~{\sim}~3~\rm {keV}$) in addition to a cut-off power law
which is typical for X-ray pulsars. However, the X-ray spectrum 
did not show the presence of cyclotron resonant scattering 
features~(CRSF) that could provide an estimate of the NS magnetic field \citep[see][]{Jaisawal18, Tao19}. 
Different methods have suggested a magnetic field of $10^{13}\rm{G}$
\citep[][]{WH18, Doroshenko18}, although \citet{Tsygankov18} suggests it could be lower.
Very recently, \citet{Zhang2019} reported results obtained from \emph{HXMT} monitoring of J0243,
finding no evidence for a cyclotron feature up to 150~keV. However, based on the spin evolution
study of J0243 performed using \emph{HXMT} data, they also suggested that the NS's magnetic field 
{\bf{is}} $\sim$~$10^{13}\rm{G}$.~However, this estimate contrasts with that proposed by
\citet[][]{Jaisawal19}, whose broad iron line (peaking at $6.67~\rm{keV}$) 
in \emph{NICER} spectra requires a dipolar magnetic field in a narrow range between $10^{11}\rm{G}$ and $10^{12}\rm{G}$
if it is to originate in the accretion disc.
The presence of a weakly magnetized neutron star is also 
supported by a sharp state transition of the timing and spectral 
properties of the source at super-Eddington accretion rates \citep[][]{Doroshenko2020}. \\

Transient X-ray pulsars are valuable natural laboratories to 
understand the evolution of magnetically-driven accretion.~In particular, details of the accretion column geometry can become clear as the mass accretion rate evolves, and J0243 is ideal for such work.~Therefore,~several X-ray studies have been undertaken with wide energy
coverage such as \emph{NICER} (WH18), 
\emph{NuSTAR} \citep{Bahramian17, Jaisawal18}, and the \emph{Hard X-ray Modulation Telescope (HXMT)} \citep{Tao19}. \\

Pulse profiles reflect the beaming pattern of X-ray emission.
WH18 studied pulse profiles of J0243 in the 0.2-100 keV energy band
using data from \emph{NICER} and \emph{Fermi}.
\citet[][]{Tsygankov18} monitored this source with \emph{Swift}-\textsc{XRT}, and 
found the pulse profiles of J0243 to change significantly 
above luminosities $\sim$~$10^{38}\rm {erg~s^{-1}}$ \citep[see][for details]{WH18,Tsygankov18}. \\

As a part of this multi-wavelength campaign, \emph{AstroSat} observed J0243 twice during its outburst.
\emph{AstroSat} is the first Indian multi-wavelength astronomical satellite \citep[][]{Agrawal06, KP14}
and was launched in 2015.~The three co-aligned X-ray instruments on-board \emph{AstroSat}:~a Soft X-ray Telescope (\textsc{SXT}) \citep[][]{KP16,KP17}, 
a Large Area Xenon Proportional Counter (\textsc{LAXPC}) \citep[][]{Yadav16,Antia17}~and a Cadmium-Zinc-Telluride Imager
(\textsc{CZTI}) \citep{Vadawale16, Bhalerao17}
give simultaneous broadband coverage from 0.3-200~keV.

In this paper, we report the results obtained from these observations. 
In \S \ref{sec:obs}, we give observational and data reduction
details, while timing analysis and results 
are presented in \S \ref{s:res_time}.
We have performed a broadband spectral analysis and the results are presented 
in \S \ref{ss:spec_analysis}.
Finally, we discuss them in \S \ref{sec:disco}.

\section{Observations \& Data Reduction}	
\label{sec:obs}

\begin{figure}
 \centering
 \includegraphics[height=3.2in, width=2.2in, angle=-90]{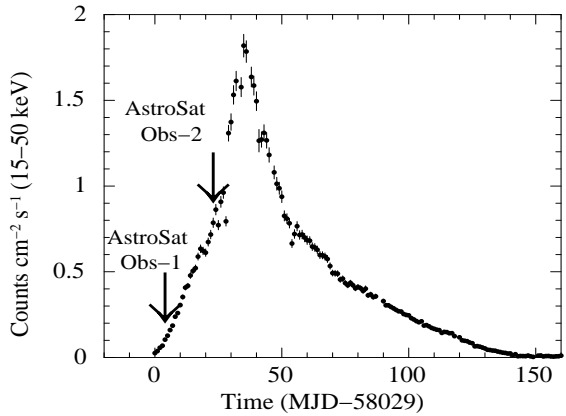}
 \caption{{\it Swift}--BAT light curve of J0243 in the 15-50 keV range, from 2017 October 3
 (MJD 58029) to 2018 February 11 (MJD 58160). Arrows mark times of {\emph AstroSat} observations, 
the first starting on MJD~58033.193, the second on MJD~58052.619.}
 \label{sw-lc}
 \end{figure}

\emph{AstroSat} observed J0243 twice as part of the Target of Opportunity~(ToO) program,
as detailed in 
Table~\ref{tab:astrosat}. 
The {\emph AstroSat} data were obtained from the ISSDC data dissemination
archive\footnote{https://astrobrowse.issdc.gov.in/astro\_archive/archive/}. 
In Figure~\ref{sw-lc},~we show the \emph{Swift}--\textsc{Burst Alert Telescope}~(\textsc{BAT}) light curve of J0243,
with 1-day binning, to indicate the epochs of the AstroSat observations.

\begin{table*}
\centering
	
	\caption{Observations made with {\emph AstroSat} during the 2017 outburst of J0243.} 
	\label{tab:astrosat}
	\scriptsize{\begin{tabular}{|c|c|l|c|c|}
	\hline        
	
		  &     & ObsID T01\_193T01\_9000001590~(Obs--1) &    \\
	\hline
		Instrument  &	Exp Time (ks) &	Start Time (UTC) & End Time (UTC)  & Background-subtracted countrate ($cs^{-1}$)\\
	\hline \hline

    \textsc{SXT}  &	 $\sim$~18   &	2017-10-07 09:19:13	& 2017-10-08 00:31:40 & $\sim$$10.62$ \\ 
    \textsc{LAXPC10} &	 $\sim$~50   &	2017-10-07 05:06:17 	& 2017-10-08 00:54:50 & $\sim$$1113$                 \\ 
  \textsc{LAXPC20}   &	 $\sim$~50   	& 2017-10-07 05:06:17	& 2017-10-08 00:54:50 & $\sim$$1118$                \\ 
   \textsc{CZTI}  &	 $\sim$~50    &	 2017-10-07 04:38:00	&2017-10-08 00:48:00  & $\sim$$0.079$                 \\ 

   \hline \hline
	
		&     & ObsID T01\_202T01\_9000001640~(Obs--2) &    \\
   
   \hline
		Instrument  &	Exp Time (ks) &	Start Time (UTC) & End Time (UTC) &  Background-subtracted Countrate~($cs^{-1}$)   \\
	\hline \hline
  \textsc{SXT}  &	 $\sim$~14   &	2017-10-26 16:10:44	& 2017-10-27 10:38:41 & $\sim$$130$ \\ 
    \textsc{LAXPC10} &	 $\sim$~46   &	2017-10-26 15:00:04 	& 2017-10-27 12:01:28 & $\sim$$9320$                 \\ 
  \textsc{LAXPC20}   &	 $\sim$~46   	& 2017-10-26 15:00:04	& 2017-10-27 12:01:28 & $\sim$$9477$                \\ 
   \textsc{CZTI}  &	 $\sim$~46    &	 2017-10-26 14:52:00	&2017-10-27 11:53:00  & $\sim$$0.742$                 \\ 
   \hline \hline

	\end{tabular}}
\end{table*}

\subsection{AstroSat Data Reduction and Analysis}

We followed standard analysis procedures for individual instruments~(\textsc{SXT}, \textsc{LAXPC}
and \textsc{CZTI}), as suggested by the instrument teams.~Data reduction
pipelines and tools disbursed by the \emph{AstroSat} Science Support Center~(ASSC)\footnote{https://astrosat-ssc.iucaa.in}
have been used for the data analysis.

\subsubsection{\textsc{SXT}}

The \textsc{SXT} instrument on-board \emph{AstroSat} is capable of
X-ray imaging and spectroscopy in the 0.3-7~keV energy range.
It has a focusing telescope and a CCD detector that was 
operated in the  
{\textquotedblleft}Fast Windowed Photon Counting"~(FW) mode
with a time resolution of 0.278~s for both observations.
In the FW mode, a fixed window 
of $10{\arcmin}{\times}10{\arcmin}$ out of the entire $40{\arcmin}{\times}40{\arcmin}$
CCD detector is used for observations \citep[see][for details]{KP16,KP17}.
We processed the \textsc{SXT} data using the {\texttt sxtpipeline}
v1.4 and the \textsc{SXT} redistribution matrices in CALDB~(V20160505).~The cleaned event files of all orbits of 
each observation were merged using \textsc{SXT} Event Merger Tool~(Julia Code
\footnote{http://www.tifr.res.in/$\sim$astrosat\_sxt/dataanalysis.html}).
The merged events files were then used to extract
images, light curves and spectra using the ftool task \texttt{XSELECT}, 
provided as part of \texttt{HEAsoft v 6.19}. \\

We checked the cleaned event files for pile-up, following the method of \citet{Sreehari19}, and
found that the J0243 observations were not affected. 
Therefore, for our analysis presented here we extracted source region files 
using a circular radius of 5.0 arcmin as indicated in the \textit{AstroSat Handbook}
\footnote{http://www.iucaa.in/~AstroSat\_handbook.pdf}. 
We have used the spectral redistribution matrix {\textquotedblleft}sxt\_pc\_mat\_g0to12.rmf'', and
the FW mode ancillary response matrix file~(sxt\_fw\_v02.arf) provided by the
instrument team for spectral analysis.

\subsubsection{\textsc{LAXPC}}

The \textsc{LAXPC} instrument on-board \emph{AstroSat} has three co-aligned
proportional counters viz., \textsc{LAXPC10}, \textsc{LAXPC20}, \textsc{LAXPC30},~each
with seven anodes arranged into five layers and each with 12 detector cells.
Due to a gain instability issue caused by gas leakage, we have not used
\textsc{LAXPC30} data.
Each \textsc{LAXPC} detector independently records the time of arrival
of each photon with a time resolution of 10~$\mu$s and works in the energy range of 3--80~keV. 
The deadtime of the \textsc{LAXPC} instrument is 43~$\mu$s. The energy resolution for
\textsc{LAXPC10}, \textsc{LAXPC20} at 30~keV is about $15{\%}$, 
$12{\%}$, respectively \citep{Yadav16, Antia17}.
\textsc{LAXPC} data collected in the Event Analysis mode~(EA) were used 
 for performing timing and spectral analysis.
 For timing analysis, we used combined data of \textsc{LAXPC10}
and \textsc{LAXPC20}.~During both observations \textsc{LAXPC10} was operating
at a lower gain.~Therefore, we have used data only from \textsc{LAXPC20} for spectral analysis. \\

Light curves and spectra were generated using the {\texttt{LaxpcSoft}} software 
package\footnote{http://www.tifr.res.in/~astrosat\_laxpc/LaxpcSoft.html}.
The background in \textsc{LAXPC} is estimated from blank sky observations,
where there are no known X-ray sources, and the count-rates are fitted 
as a function of latitude and longitude to provide the background estimate 
for J0243.
The soft and medium energy X-rays
do not reach the bottom detector layers. Therefore, to avoid
additional background, we extracted light curves using data from the top layer
for energies up to 25~keV while for the hard energy bands and the 
average energy band~(3-80 keV), we have used data from all layers of each detector.
We have used response files generated by software to obtain channel to
energy conversion information while performing the energy filtering.
For the extraction of source spectra, a similar procedure is followed
except for background counts being averaged
over the duration of the observation.
\subsubsection{\textsc{CZTI}}
The CZTI instrument \citep{Bhalerao17} on-board \emph{AstroSat}
is a 2-D coded mask imager with solid state
pixelated CdZnTe detectors.
\textsc{CZTI} has an
energy range of 20--200~keV, providing an angular resolution
of 8~arcmin with a field of view of $4.6^{\circ}{\times}4.6^{\circ}$.
Events recorded by \textsc{CZTI} are time-stamped with a resolution
of $20~{\mu}s$. \\

To analyse the \textsc{CZTI} data, we have used the 
{\it Level~2 pipeline} and followed the procedures indicated in its 
user guide
\footnote{http://astrosat-ssc.iucaa.in/uploads/czti/
CZTI\_level2\_software\_userguide\_V2.1.pdf}.
This analysis software generates spectral response specific to a given observation.
Simultaneous measurement of background
is available from the coded mask imager from which the
background-subtracted products were generated.~We have used the data from \textsc{CZTI} only for performing a timing study and it is not included in the spectral analysis presented in \S~\ref{ss:spec_analysis}.

 \begin{figure*}
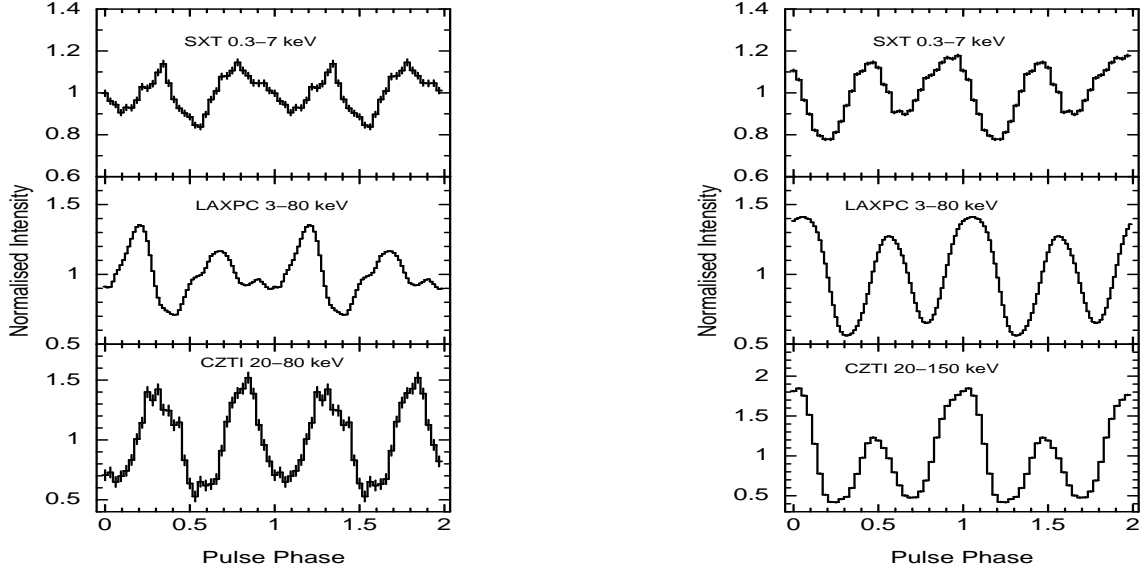

\begin{minipage}{0.45\textwidth}
\includegraphics[height=\columnwidth,width=0.85\columnwidth]{fig4.eps}
\end{minipage}
\hspace{0.05\linewidth}
\begin{minipage}{0.45\textwidth}
 \includegraphics[height=\columnwidth,width=0.85\columnwidth]{fig5.ps}
\end{minipage}

\caption{Average, background-subtracted pulse profiles of J0243 obtained with (top to bottom)
\emph{AstroSat}'s {\textsc{SXT}}, {\textsc{LAXPC}} and \textsc{CZTI} for Obs--1~(left)
	 and for Obs--2~(right).~Error bars represent 1$\sigma$ uncertainties.}
\label{Avg-PP}
\end{figure*}

\begin{table*}

  \begin{minipage}{0.5\textwidth}
	
	  \caption{\emph{AstroSat} pulse period measurements of J0243 during the 2017 outburst} 
	\label{tab:pulseperiod}
	\scriptsize{\begin{tabular}{|c|c|l|c|c|}
	\hline        
	
	          &      Obs--1 &    \\
	\hline
		Instrument & Energy Band~(keV) &	Pulse Period~(s) \\
	\hline 

    \textsc{SXT}  & 0.3--7	  & $9.8530\pm0.0004$            \\ 
    \textsc{LAXPC} & 3-80	  &  $9.8532\pm0.0003$                \\ 
    \textsc{CZTI}  & 20--80	  & $9.8527\pm0.0003$         \\ 

   \hline 
	
   &      Obs--2 &             \\
   
   \hline
		Instrument  & Energy Band	&Pulse Period (s)    \\
	\hline 
  \textsc{SXT}  &	0.3--7  & $9.8505\pm0.0003$              \\ 
    \textsc{LAXPC} &	3-80   & $9.8502\pm0.0002$              \\ 
   \textsc{CZTI}    &   20--150  &  $9.8500\pm0.0003$               \\ 
  \hline 
	\end{tabular}}
 \flushleft

  \end{minipage}

\end{table*}

\section{Timing Analysis and Results}
\label{s:res_time}
\subsection{Average Pulse Profiles}
\label{avg-pp}
The source and background light curves with a binsize of 300~ms
were extracted in the 0.3--7~keV energy band, using the \textsc{SXT} data.
For \textsc{LAXPC} and \textsc{CZTI}, the source and background light curves
were extracted with a binsize of 100~ms using data in the 3--80~keV
and 20--200~keV band, respectively.
The average count rates observed in each instrument are given in Table~\ref{tab:astrosat}.
Barycentric correction was applied to the background subtracted light curves using
tool {\texttt {as1bary}}\footnote{http://astrosat-ssc.iucaa.in/?q=data\_and\_analysis}.
This is a modified version of the well-known \texttt{AXBARY} task of the \texttt{HEAsoft} package.~To search
for pulsations in the light curves obtained from all three instruments, we have used the standard $\chi^2$ maximization technique \citep{Leahy87}
applied the \texttt{efsearch} task of FTOOLS~(see Table~\ref{tab:pulseperiod}). \\

Independent methods such as CLEAN \citep{Roberts87} and the Lomb-Scargle periodogram \citep{Lomb1976,Scargle1982,Horne1986},
as implemented in the PERIOD program distributed with the \textsc{Starlink Software Collection}\footnote{http://starlink.eao.hawaii.edu/starlink}
\citep{Currie14} were also used to estimate the spin period.
We obtained consistent values of spin period from \textsc{SXT} and \textsc{LAXPC} data using these methods.
However,~a different period ($\sim$9.15s) was found in the 20--200~keV band \textsc{CZTI} light curve of Obs--1.
On investigating, we found that the background
dominates at higher energies in this observation, and so we restricted the \textsc{CZTI} light curve to
the 20--80~keV band in our analysis.~Background domination above 150~keV was also seen in the
\textsc{CZTI} light curve of Obs--2,~and so we restricted the energy band here to 20--150~keV
in creating our pulse profiles.~We also calculated the False Alarm Probability~(FAP) \citep{Horne1986} 
so as to compute the periodogram's peak power significance, and it was found to be above 95$\%$,
except for the 20--80~keV \textsc{CZTI} light curves of Obs~1.~We obtained a large FAP value for the \textsc{CZTI} light curves of Obs~1,
suggesting that the detected peak may not be significant.
The uncertainity estimated by these methods corresponds to the minimum error on the period.
A reliable estimate of the error can be obtained using simulation of a large number 
of light curves via Monte Carlo or randomization methods \citep[see e.g.,][]{Boldin13}.
We found that the periods estimated by all these methods are consistent with each other, as given in Table~\ref{tab:pulseperiod}. \\

Figure~\ref{Avg-PP} shows the average pulse profiles during both observations. These are obtained by folding
the \textsc{SXT}, \textsc{LAXPC} and \textsc{CZTI} light curves on the period determined
from the \textsc{LAXPC} high-time resolution data,
and each having 32,~64 and 32 phase bins, respectively. 
From Figure~\ref{Avg-PP}, it is clear that during Obs--1 the \textsc{SXT}
pulse profile shows double-peaked structure between phases 0.2-0.4 and 0.6-1.0.
There is a significant evolution in the pulse profile
shape in the \textsc{LAXPC} energy band. The 3--80~keV pulse profile
is relatively more complex compared to the \textsc{SXT} profiles. There exists a double-peaked behaviour
followed by some structure in
the rest of the profile.~In the \textsc{CZTI} band~(20--80~keV), the pulse profile changed to a shape similar to that observed with the \textsc{SXT}. 
Pulse profiles during Obs--2
are relatively simpler compared to the first one. A double-peaked behaviour is observed 
in both \textsc{SXT} and \textsc{LAXPC} light curves. 
The \textsc{CZTI} profile also shows a double-peaked structure, but at slightly
different phases of 0-0.2 and 0.4-0.6.

\subsection{Energy-resolved Pulse Profiles}
\begin{figure*}
\centering
\includegraphics[height=1.75\columnwidth, width=\columnwidth, angle=-90]{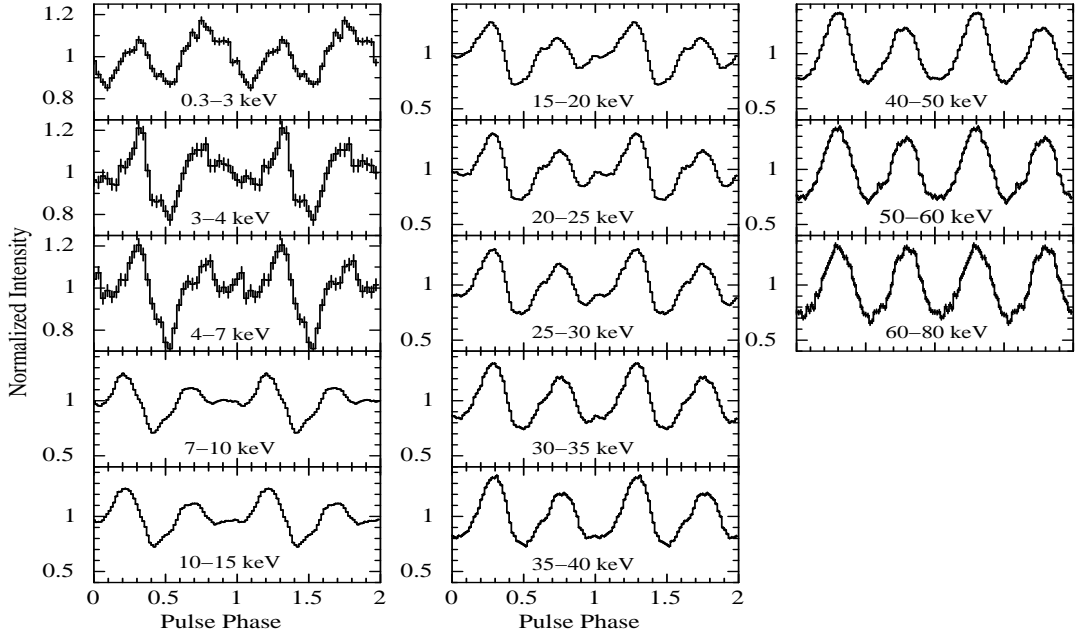}
\caption{Energy-resolved pulse profiles of J0243 obtained from the background-subtracted light 
	 curves of {\textsc{SXT}}~(0.3-7~keV) and {\textsc{LAXPC}}~(7.0-80.0~keV) during Obs--1.~The error bars 
	 in each panel represent 1$\sigma$ uncertainties.~Two cycles in each panel 
are shown for clarity.}   
\label{erpp1}
\end{figure*}

\begin{figure*}
\centering
\includegraphics[height=2\columnwidth, width=\columnwidth, angle=-90]{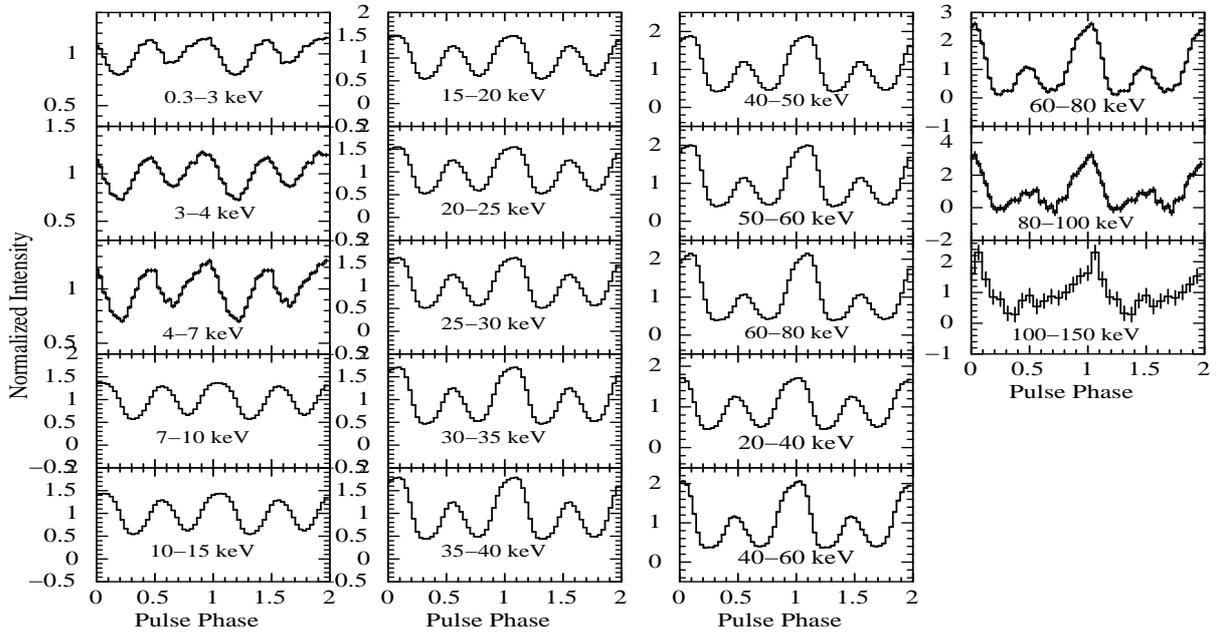}
\caption{Pulse profiles of J0243 obtained from the background-subtracted light 
curves of {\textsc{SXT}}~(0.3-7~keV), {\textsc{LAXPC}}~(7.0-80.0~keV) and {\textsc{CZTI}}~(20-150~keV) during Obs--2.
These profiles also show strong energy dependence and the double-peaked 
profile evolves with energy. Pulsations 
were clearly detected in the light curves up to 150 keV.~The error bars 
	in each panel represent 1$\sigma$ uncertainties. Two cycles in each panel 
are shown for clarity.}   
\label{erpp2}
\end{figure*}

We note that the pulse profiles obtained across the three instruments
show prominent differences, indicating 
a strong energy dependence. 
Energy-resolved pulse profiles are also an important tool to investigate the pulsar emission geometry.
We extracted light curves in narrow energy bands using data from \textsc{SXT}, \textsc{LAXPC} and \textsc{CZTI} during both observations of the pulsar. Light curves in 0.3--3~keV, 3--4~keV and 4-7~keV range were extracted for \textsc{SXT}. A total of 10 energy bands~(7--10~keV,~10--15~keV,~15-20~keV,~20--25~keV,~25--30~keV,~30--35~keV,~35--40~keV,~40--50~keV,~50--60~keV,~60--80~keV) were chosen to extract light curves for \textsc{LAXPC} and six energy ranges (20--40~keV, 40--60~keV, 60--80~keV, 80--100~keV, 100--150~keV and 150--200~keV) were selected for \textsc{CZTI}.
To verify the detectability of the pulsations in individual bands,~we performed period search on these energy-resolved light curves and computed significance of
detection using the same methods as for the average light curves (see Section~\ref{avg-pp}).
We found that false alarm probabilities lie between
0.00 and 0.01 with 95$\%$ confidence for all the \textsc{SXT}
and \textsc{LAXPC} energy-resolved light curves.~In the case of Obs--1, we obtained large values
of FAP for all energy-resolved \textsc{CZTI} light curves, indicating that the period detection is not significant.
Moreover, a different period value was found for the 150--200~keV light curve of Obs--2, suggesting that it
is dominated by the background.~Therefore, we did not include \textsc{CZTI} data of Obs--1 and the
150--200~keV light curve of Obs--2 in performing energy-resolved analysis.
Figures~\ref{erpp1} and \ref{erpp2} show the background subtracted, energy-resolved pulse profiles obtained from Obs--1 and Obs--2 of J0243, respectively.{\footnote{To verify the internal consistency of these data, we checked the pulse profiles obtained in the two energy bands of 3--7~keV (where \textsc{SXT} and \textsc{LAXPC} overlap) and 20--80~keV (where \textsc{LAXPC} and \textsc{CZTI} overlap) and they are effectively identical.}} \\

\begin{figure*}
\centering
\begin{minipage}{0.45\textwidth}
\includegraphics[height=\columnwidth,width=\columnwidth, angle=0]{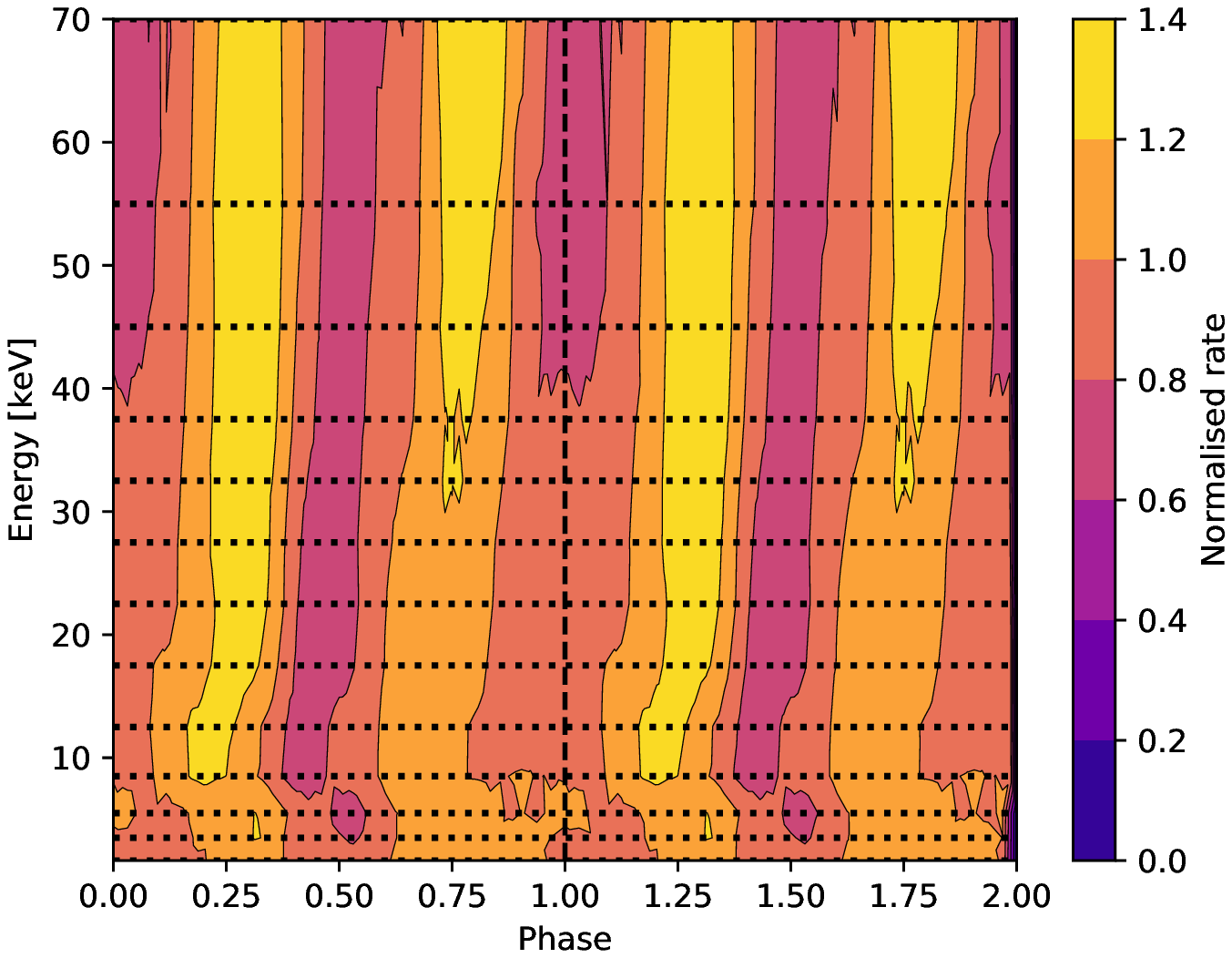}
\end{minipage}
\hspace{0.05\linewidth}
\begin{minipage}{0.45\textwidth}
 \includegraphics[height=\columnwidth,width=\columnwidth,angle=0]{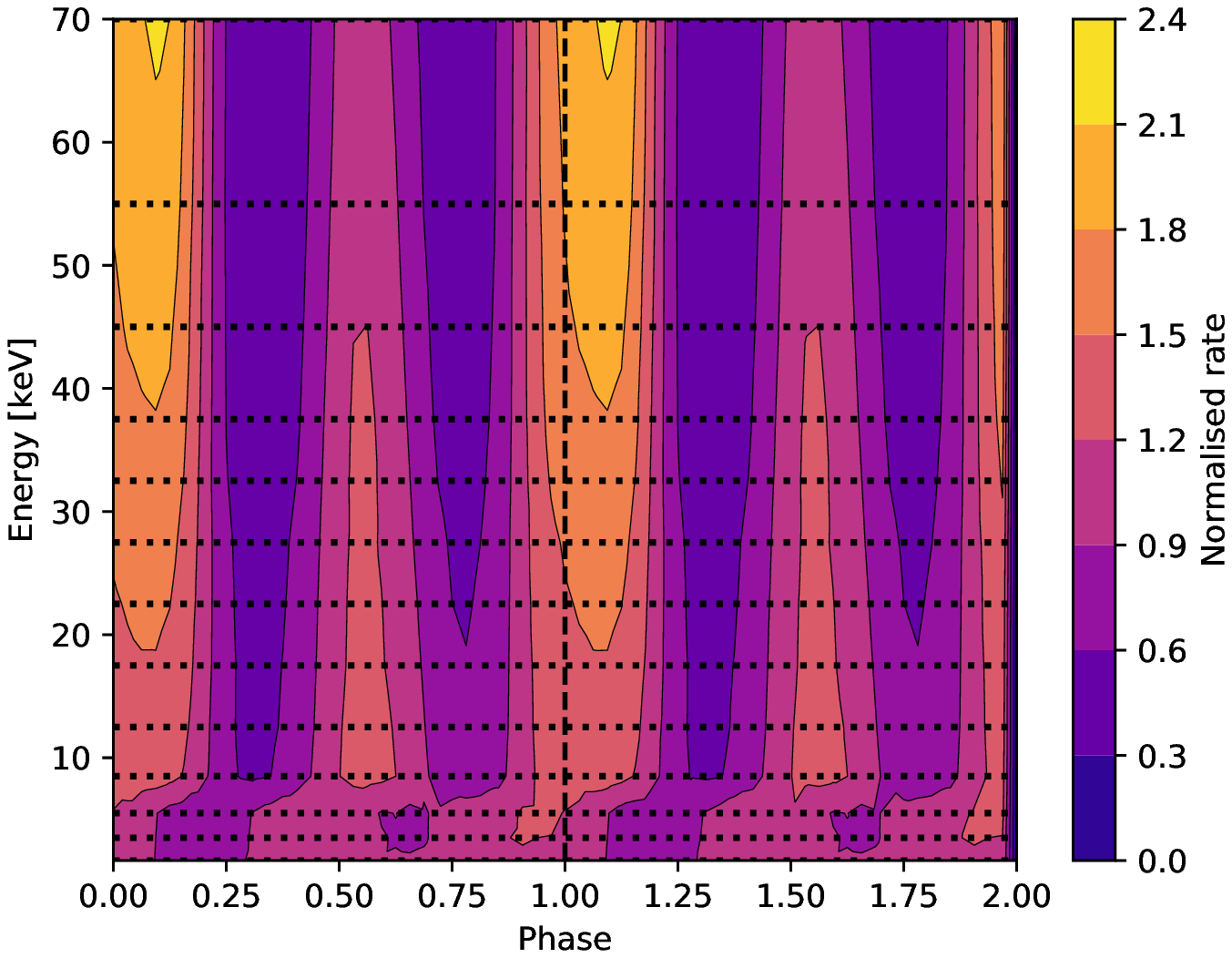}
\end{minipage}
\caption{Energy-resolved pulse profiles for J0243 from \textsc{SXT} and \textsc{LAXPC} data 
of Obs--1~(Left) and Obs--2~(Right).~Yellow~(Purple) colour correspond to higher~(lower) count rates
at different phases and energies.~Horizontal cuts are pulse profiles for the selected energy
range while the vertical cut represent spectra during one complete rotation cycle.}
\label{Heatmap}
\end{figure*}

It is evident from Figure~\ref{erpp1} that the pulse profiles are double-peaked and evolve with increasing energy.
The profiles below 7~keV are quite complex.~In addition to  
twin peaks, there exist several structures at other pulse phases.
At energies above 7~keV, emergence of a secondary peak makes the profile 
asymmetric with primary and secondary peaks appearing at pulse phases 0.2 and 0.7, respectively.
We observed that above 60~keV, the pulse shape changes to a symmetric double-peaked
profile with two peaks at almost the same intensity.
The predominance of these structures is also confirmed by the analysis
carried out in the left panel of Figure~\ref{Heatmap}.
Here we represent with yellow~(purple) colors the phases of all the pulse profiles
of \textsc{SXT} and \textsc{LAXPC} shown in Figure~\ref{erpp1} where the source intensity
is higher~(lower). \\

For Obs--2, we found that the pulse profiles
are simpler at lower energies~(below~7~keV) compared to the previous observation.
There exist two comparable peaks in the profile 
at energies below 10 keV.~However, above 10~keV the pulse profile changes from symmetric to asymmetric,
having a primary peak at around phase 0.0 and secondary peak at around phase 0.5.~A similar profile was also 
observed with \emph{CZTI} (see  Figure~\ref{erpp2}).~This behaviour of the pulse profiles 
is also evident from the right panel of Figure~\ref{Heatmap}.
Similar energy dependence of pulse profiles was seen in \emph{NICER} and \emph{Fermi}-\textsc{GBM}
data taken close to the dates of our \emph{AstroSat} observations (see WH18). \\

\begin{figure*}
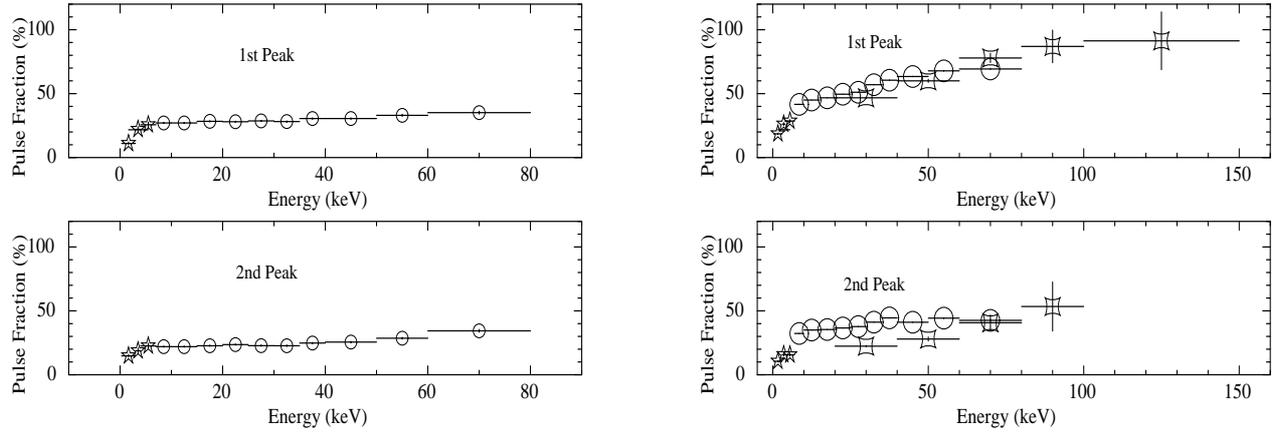

\centering
\begin{minipage}{0.45\textwidth}
\includegraphics[height=\columnwidth,width=0.75\columnwidth, angle=-90]{fig8.eps}
\end{minipage}
\hspace{0.05\linewidth}
\begin{minipage}{0.45\textwidth}
 \includegraphics[height=\columnwidth,width=0.75\columnwidth,angle=-90]{fig9.eps}
\end{minipage}
\caption{Pulse fraction variation of J0243 as a function of energy 
for Obs--1~(Left) and Obs--2~(Right).
The horizontal bars represent the energy bins used.}
\label{PF}
\end{figure*}

In Figure~\ref{PF}, we show the dependence of pulse fraction~(PF) on energy.~The PF is defined
as the ratio between the difference of maximum~($I_{max}$) and minimum~($I_{min}$) intensity to their sum:
($(I_{max}-I_{min})/(I_{max}+I_{min})$) and allows 
us to estimate the fraction of photons contributing to observed pulsations.
A careful examination of the PF indicates that during Obs--1, the PF increases with energy for both the peaks
observed in the pulse profile.~The PF for first peak~(around 0.2 pulse phase) increased from $\sim$~$11\pm1{\%}$~(1.65~keV) to~$35\pm1{\%}$~(70~keV) while for 
the second peak~(around 0.7 pulse phase) it 
increased from $\sim$$~15\pm1{\%}$~(1.65~keV) to
$34\pm1{\%}$~(70~keV).~For Obs--2, we observe that the first peak~(around~0.0)
shows an increase in the PF with increase in energy~(from $\sim$~$18.5\pm0.5{\%}$~(1.65~keV) to~$78\pm3{\%}$~(70~keV)), while for the second peak~(around 0.5)
it increased from $\sim$~$10.6\pm0.5{\%}$~(1.65~keV) to $43\pm1{\%}$~(70~keV).

\section{Spectral Analysis and Results}
\label{ss:spec_analysis}
We have performed spectral fitting using \texttt{XSPEC} v-12.9.0~\citep{Arnaud96}.
The extracted \textsc{SXT} and \textsc{LAXPC} spectra were grouped using the FTOOLS task 
{\textquoteleft}\texttt{grppha}' to have a minimum of 25 counts per bin.
We have used the 1-7~keV energy range of \textsc{SXT} and ignored data below 1~keV to avoid systematic uncertainties
in the calibration at very low energies.~For Obs--1, we found that the \textsc{LAXPC} background dominates at higher energies (see Appendix~A)
and, in order to avoid an undesirable contribution from instrumental systematics, we considered only the
4-25 keV energy range when performing spectral fitting.~For Obs--2 the energy range of 4-70~keV 
was used.~To include the effects of Galactic absorption we used the
{\textquoteleft}\textit{tbabs}' model component 
with abundances from \citet{Wilms00} and cross-sections as given by
\citet{Verner96}.~A multiplicative term~({\it CONSTANT}) was
added to the model to account for calibration uncertainties between
\textsc{SXT} and \textsc{LAXPC}.~This factor was
fixed to 1 for the \textsc{SXT} data and was allowed to vary for \textsc{LAXPC20}. \\ 

Accretion-powered X-ray pulsars radiate powerfully over a wide energy range, from thermal seed photons at soft X-rays to the reprocessed emission (inverse Comptonization of thermal seed photons) at hard X-rays.
Their broadband continuum spectra are typically described with
a combination of a black-body (for the low energy excess) and a power law with quasi-exponential high energy
cut-offs of various forms.~One of the most widely used continuum models has a high energy exponential cut-off \citep[see e.g.,][]{White83,Mihara95,Coburn02,Furst14}. \\

The continuum emission of J0243 obtained with \emph{NuSTAR}
and \emph{HXMT} was studied using an absorbed black-body ({\texttt bbodyrad})
and a cut-off power law ({\texttt cutoffpl}) model \citep{Bahramian17, Jaisawal18,Zhang2019}.
There also exist other phenomenological \texttt{XSPEC} models such as high energy cut-off power law
({\textquoteleft}{\texttt highecut}'), a combination of two negative and positive power laws with
exponential cutoff ({\textquoteleft}{\texttt {NPEX}}').
Other local models such as power law with Fermi-Dirac cut-off \citep[{\textquoteleft}{\texttt fdcut';}][]{Tanaka86}
and a smooth  high  energy  cut-off  model \citep[{\textquoteleft}{\texttt newhcut';}][]{Burderi00}
are also often used to study the spectra of accretion-powered pulsars.
We tried to model the continuum emission 
of J0243 using all of them (see Tables~\ref{spec-para-obs1},\ref{spec-para-obs2}).
The \textsc{SXT} spectra were corrected for gain offset
during the observations using the \texttt{gain fit}
command with fixed slope of 1.0 and best fit offset of $\sim$~0.09~eV.
An offset correction of 0.03--0.09 keV is needed in quite a few \textsc{SXT} observations.

The following two models provided better fits to the J0243 spectra of Obs--1, in terms of absence 
of systematic residuals and smaller $\chi^{2}_{\nu}$
values:~{\texttt tbabs*(highecut*powerlaw+bbodyrad)},
 {\texttt tbabs*(newhcut*powerlaw+bbodyrad)},~where
{\textquoteleft}\texttt{newhcut}' is the modified version of the
 high energy cut-off model, smoothed around
 the cut-off energy.~The mathematical form
 of this model can be found in \citet{Burderi00}.
 The constants in this model are calculated 
 internally assuming the continuity of the
 intensity function and its derivative in the range of
 $E_C\pm{\bigtriangleup}E$.~We fixed ${\bigtriangleup}E$
 to 5.0~keV while performing the spectral fitting.
 This was done for consistency with the method adopted in the spectral
 study of accretion powered pulsars using this model \citep[see e.g.,][]{Jaisawal15,Maitra17}.
We added a systematic error of 1\% over the entire 1-25.0 keV energy band.
In Table~\ref{spec-para-obs1}, we give the best-fit parameters
and 90~$\%$ confidence ranges obtained with these models. \\
The presence of a neutral iron $K_{\alpha}$ line at 6.4~keV was detected
in \emph{NuSTAR} spectra of J0243 \citep{Jaisawal18, Tao19}.
Therefore, we tried adding a Gaussian component to the above-mentioned best-fit
continuum models, however, we found that a neutral iron $K_{\alpha}$ line
is statistically not required.  \\

\begin{figure*}
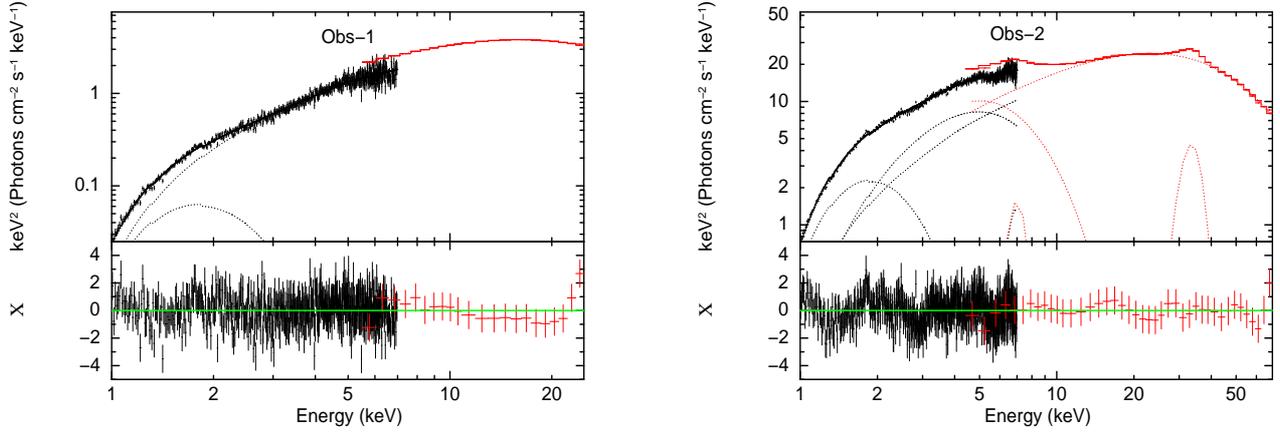

 \centering
\begin{minipage}{0.45\textwidth}
\includegraphics[height=\columnwidth,width=0.75\columnwidth, angle=-90]{fig10.ps}
\end{minipage}
\hspace{0.05\linewidth}
\begin{minipage}{0.45\textwidth}
 \includegraphics[height=\columnwidth,width=0.75\columnwidth,angle=-90]{fig11.ps}
\end{minipage}
	 \caption{\emph{AstroSat} spectral data from J0243 and model component fits to detectors~{\textsc SXT},
	 {\textsc LAXPC} for Obs--1~(1--25~keV; Left) and Obs--2~(1--70.~keV; Right).~The 30~keV Gaussian is due to the calibration Xenon Edge.
	 These unfolded spectra are plotted using {\it{eeuf}} in {\texttt{XSPEC}}, corresponding to units of ${\nu}f_{\nu}$.~The best-fit obtained using Model~1
	 of Obs--1 and Obs--2 are shown.}
 \label{spec2}
 \end{figure*}

The same continuum spectral models as above were first tried to
fit the spectral data of Obs--2.
However, we found that the four component models
used in the first observation were
inadequate to give a good~(or acceptable) fit to the data.
Residuals around 30 keV due to the Xenon calibration edge (Antia et al. 2017) were observed in the \textsc{LAXPC20} spectra
and were modeled using a Gaussian.~A similar feature has also been found in the \textsc{LAXPC} spectra
of other sources \citep[see e.g.,][]{Sharma2020,Banerjee2020}.~We also added a larger systematic ($3{\%}$) value 
to account for uncertainties in response calibration over such a wide energy band 4-70~keV \citep[see e.g.,][]{Mudambi2020}.
~Additional systematic residuals at low energies were also observed, therefore,
following \citet{Tao19} we added an additional black-body component~(\texttt{bbodyrad})
to the spectra.~This black-body component may be associated with the thermal emission
from the photosphere of optically-thick outflows or from the extended accretion
column as proposed for the case of super-Eddington accretion.~We observed that reasonably
good fits were obtained by addition of this component to the model~(see Figure~\ref{spec2}).
We also found that spectral fits required an additional Gaussian component to fit the 
iron emission feature observed at around 6.9~keV.
As it was difficult to constrain all the line parameters, therefore, we fixed the
line width to 0.5~keV.
The best-fit parameters are given in Table~\ref{spec-para-obs2}.
From this table, we can see that~Model~1 {\texttt tbabs*(highecut*powerlaw+bbodyrad+bbodyrad \\
+gaussian)}
describes the broad-band spectra well and, therefore, to compute the significance of the detected iron emission
line, we have used Model~1.
We found that on adding this additional Gaussian component at around 6.9~keV, the value of
$\chi^{2}$ decreased from 643(628) to 615(626) for 2 degrees of freedom, corresponding to an F value of
13.8, and an F-test false alarm probability of $1.3\times10^{-6}$. This suggests that the detection of an iron
line feature is statistically significant. \\

To further investigate the significance of this emission feature
we simulated 10,000 spectra assuming the {\texttt tbabs*(highecut*powerlaw+bbodyrad+bbodyrad)}
model to be true.~We searched for the presence of an iron line in each of these
data sets by comparing the best fit $\chi^{2}$ values
with and without a Gaussian component.
We then compared the F statistic of each simulation against that for the observed data.
Finally, we infer the probability of a chance improvement
of $\chi^{2}$ by counting how many times the simulated values
of F were larger than obtained from the real data. In all cases, the estimated
chance probability was lower than observed in the real data, implying a significance of
$3.9~{\sigma}$.

\section{Discussion}
\label{sec:disco}

\emph{AstroSat} observed J0243 twice during its 2017-18 outburst, and we have analysed data obtained over a broad energy range (0.3-150 keV) with all three of its X-ray instruments.~In Section~\ref{sec:pp} we discuss the results of our timing study, while Section~\ref{sec:spectroscopy} addresses the spectroscopic results.

\subsection{Evolution of the Pulse Profiles}
\label{sec:pp}

J0243 was accreting at sub-Eddington level during Obs--1, while during Obs--2 the pulsar was super-Eddington. We probed the light curves extracted in narrow energy bands and found that significant pulsations were detected up to 150~keV during Obs--2. However, for the relatively fainter observation (Obs--1), pulsations were detected only up to 80~keV.~The average pulse profiles
revealed a double-peaked behaviour during both observations,~separated by $\sim$~19~days.
The existence of two peaks in the pulse profiles can be due to the contribution from both magnetic 
poles of the neutron star or two sides of a fan beam from one pole.
The pulse profiles created using data from each instrument showed a strong energy dependence.
During Obs--1 the soft energy pulse profiles 
are quite complex compared to higher energies, while for Obs--2, the pulse profiles
in all energy bands are relatively simpler, but the modulation is much larger at higher energies.
Differences seen in the pulse profiles of these two observations could be due to changes in their accretion levels.
WH18 also observed similar dependence of the pulse profiles on the X-ray luminosity 
using data from \emph{NICER} and \emph{FERMI}-\textsc{GBM},
however, the \emph{AstroSat} data allowed us to probe into these profiles up to 150~keV using narrow energy bands.
Luminosity dependence of the pulse profiles  
has also been observed in several other X-ray pulsars \citep[see e.g.,][and references therein]{White83,Nagase89,Doroshenko2020}. \\

During these observations~(Obs--1 \& Obs--2), we found
that the PF of both peaks in the pulse profiles showed an increase with energy.
This indicates that the higher energy photons contribute to the X-ray pulsations.
\citet{Tao19} performed a PF evolution study using the 5 \emph{NuSTAR} observations, revealing that the PF
increased with increasing
energy when J0243 was super-Eddington.
They suggested that the cut-off power law dominates towards higher energies,
and is responsible for the associated increase in PF, as also observed in our pulse profiles
of Obs--2.
We note that the pulse profiles of NGC~300~ULX1 also showed similar high values
of PF \citep[see][]{Carpano18}.
Moreover, during the 2016 super-Eddington outburst of SMC~X--3 
\citep{Townsend17} a smooth increase in the PF with energy was observed \citep{Tsygankov17a},
similar to that observed during Obs--2 when J0243 was accreting at a super-Eddington level.
In a few X-ray pulsars it has been found 
that close to the cyclotron line energies PF shows a non-monotonic dependence on energy
\citep[see e.g.,][]{Tsygankov07}. Thus, the smooth behaviour observed is consistent with the 
fact that we do not observe any strong features
(e.g., CRSF) in the source energy spectrum.

\subsection{Broadband Spectroscopy}
\label{sec:spectroscopy}
The best fit to the spectra of Obs--1 was obtained using
the following model:~\texttt {tbabs*(highecut*powerlaw+bbodyrad)} 
while for Obs--2 we required two additional model components:~a hot $\sim$~1.2~keV black-body and a Gaussian ($\sim$6.9~keV) component to obtain the best fit.
Assuming a distance of 7~kpc, the unabsorbed X-ray flux measured during the first and second observations translates 
to $L_{X}$ of $7.4\times10^{37}$
and $6.3\times10^{38} \rm erg~s^{-1}$ in the 1-70~keV band, respectively.
This indicates that during the first \emph{AstroSat} observation the source was
accreting at sub-Eddington level, increasing to super-Eddington during the second observation.~\citet{Becker12} and \citet{Mushtukov15} calculated the critical luminosity~($L_c$)
of a neutron star
which marks the transition between the coulomb-dominated and radiation-dominated
accretion flow.~Assuming canonical neutron star parameters they found that 
$L_c$ is of the order of $10^{37}~\rm{erg~s^{-1}}$.~However, results obtained using 
\emph{NICER} and \emph{Fermi}-\textsc{GBM} suggested that J0243 has a much higher value of
$L_c$ of the order of $\sim$~$10^{38} \rm{erg~s^{-1}}$ (for details see WH18).
Thus, it seems that during Obs--1 the source was in its sub-critical accretion regime
while during Obs--2 it was super-critical.
This is also evident from prominent changes observed in the pulse profiles of J0243 during Obs--1 and 2
(see Section~\ref{sec:pp}). \\

During Obs--1, we observed the black-body temperature to be 0.3-0.4~keV, arising from a radius of about $25-38$~\rm{km}.
For Obs--2 the two values of temperature found are $\sim$~0.4~keV and 1.2~keV and the estimated values of black-body radius are
about $121-142$~\rm{km} and $18-19$~\rm{km}, respectively.
\citet{Tao19} studied the spectra of J0243 using \emph{NuSTAR}
and observed a blackbody temperature of about 2--3~keV during the sub-Eddington accretion level.
This thermal emission is thought to be arising from the hot spot of a
neutron star which gets hotter~(4.5~keV) during the super-Eddington phase, with \citet{Tao19}
suggesting that two additional black-body components with temperatures of about 1.5~keV and 0.5~keV 
are needed during the super-Eddington accretion level.
Based on the radius measurements, the origin of these additional black-body components is suggested to be the top of the accretion column 
and optically thick outflows, respectively.
Therefore, it may be possible that the thermal emission 
observed at a temperature of about 1.2~keV during Obs--2 is due to the emission from the accretion column
while the origin of the lower temperature~($0.38$~\rm{keV}) 
black-body component is due to the possible presence of optically-thick outflows. \\
However, we note that a recent study \citet{Jaisawal19}
suggested that, in the ultra-luminous state, the iron line is complex, and if accurately
modelled, then the 2 additional black-body components are not required at extreme luminosity.

\citet{Mushtukov17} proposed that during super-Eddington accretion, the presence of an accretion envelope plays a key role in 
the accretion process at extreme mass accretion rates.~It is expected to significantly modify
the timing and spectral properties of ULPs, with smoother more sinusoidal pulse profiles observed and 
a softer X-ray spectrum due to the reprocessing of the photons emitted from near the neutron star
by the optically-thick accretion envelope. 
J0243 exhibits complex pulse profiles at lower energies, and 
the pulsed emission is observed up to 150keV. 
The observed black-body temperature is also much lower than what is expected for 
the accretion envelopes around ULPS (${\geq}$~1~\rm{keV}). 
The absence of signatures for reprocessing the central emission in J0243 may,
however,~be attributed to its lower accretion rate than in the classical ULPS ($L_X {\sim} 10^{40} \rm{erg~s^{-1}}$){\bf{,}}
in which case the opacity of the accretion envelope is not enough to reprocess most 
of the emission from the central compact object. Interestingly, other X-ray pulsars with 
very high accretion rates;~SMC~X--3 \citep{Tsygankov17a} and NGC~300~~ULX1 \citep{Carpano18} albeit with lower 
$L_X$ than the classical ULPs, also  exhibit complex pulse profiles and high pulsed fractions. 
These sources might therefore act as an important connecting bridge between 
the classical X-ray pulsars and ULPs. \\

\citet{Jaisawal19} found a narrow 6.42~keV line when the source was in the sub-Eddington regime.
The absence of an iron emission feature in the \textsc{LAXPC} spectra
(when the source was accreting at a sub-critical level) could be due to the
limited energy resolution of the instrument, which is $\sim$$20\%$ at 6.4~keV
\citep[see][]{Yadav17}.~As an example, \citet{Sharma2020} did not find any residuals around 6.4~keV
in the combined spectra of \textsc{SXT} and \textsc{LAXPC}, while systematic residuals were seen
in the simultaneously observed \emph{XMM-Newton} spectra.
As the pulsar luminosity approaches the
Eddington limit (${\sim}1.8{\times}10^{38}{\rm{ergs^{-1}}}$ for a $1.4M_{\odot}$ neutron  star),
the iron line broadens{\bf{,}} with significant contributions from 6.67 keV (Fe~XXV), 
and 6.97 keV (Fe~XXVI) features \citep[see][for details]{Jaisawal19}.
Thus, this might be
the reason that we observed an emission line feature at around 6.9~keV
with \textsc{LAXPC} during Obs--2.

\begin{table*}
	\caption{Spectral fit parameters of Obs--1 with phenomenological models (1--25~keV). }
\label{spec-para-obs1}
\begin{tabular}{ c c c c c c c c c}
\hline
\hline
Parameters & Model 1 & Model 2 &  Model 3 & Model 4  & Model 5   \\
\hline
&   &  Obs--1~(T01\_193T01\_9000001590)  &  \\
\hline
 $N\rm{_{H}}$ ($10^{22}\rm{cm^{-2}}$) &  $0.93\pm0.05$ & $1.10\pm{-0.07}$ &  $1.03\pm0.06$  &  $0.97\pm0.06$     & $1.21\pm0.07$  \\ [0.1cm]
 
$\Gamma$  & $0.58\pm0.04$  & $0.45\pm0.03$  &  $0.25\pm0.03$  &  $0.63\pm0.05$  &    $0.69\pm0.03$              \\ [0.1cm]

	$E_{cut}$ (keV)  &  $4.8\pm0.2 $ &  $10.2\pm{0.3}$  &  $7.1\pm0.4$  & $5.36_{-0.22}^{+0.21}$  &  $1~(fixed)$      \\ [0.1cm]

$E_{fold}$ (keV) &  $11.3\pm0.4 $  & -  &-  &  $11.6_{-0.10}^{+0.05}$   &   $9.8\pm0.2$                                           \\ [0.1cm]

$kT\rm_{bb}~(keV)$  & $0.33\pm0.02$ & $0.26\pm0.01$  & $0.28\pm0.01$   & $0.30\pm0.01$   & $0.23\pm0.01$                        \\ [0.1cm]

$N^{a}$   & $0.142\pm0.007$ & $0.178\pm0.008$ & $0.158\pm0.006$  &     $0.153\pm0.01$     &      $0.39\pm0.02$                \\ [0.1cm]

$const_{LAXPC}$  & $1.20\pm0.02$   & $1.20\pm0.02$   &  $1.20\pm0.02$  & $1.20\pm0.02$     & $1.20\pm0.02$                       \\ [0.1cm]
${Unabsorbed Flux (1-25 keV)}^b$  & $1.02\pm0.04$ & $1.02\pm0.04$ & $1.02\pm0.04$ & $1.02\pm0.04$ & $1.02\pm0.04$  \\ [0.1cm]
${Unabsorbed Flux (1-70 keV)}^b$  & $1.26\pm0.04$ & $1.26\pm0.04$ & $1.26\pm0.04$ & $1.26\pm0.04$ & $1.26\pm0.04$  \\ [0.1cm]
Reduced ${\chi}^2$~(dof)          & 1.13~(580)  &  1.36~(581)  & 1.26~(580)  & 1.19~(580)  &        1.52~(580)                   \\ [0.1cm]
\hline

\end{tabular}
\\
\begin{flushleft}
{{\bf{Note}}:  
{a $\rightarrow$ Normalization~($N_{}$)
     is in units of $\rm{photons~cm}^{-2}~\rm{s}^{-1}~\rm{keV}^{-1}$ at 1~keV.} \\
{b $\rightarrow$ Unabsorbed flux in units $10^{-8} ergs \, cm ^{-2} s^{-1}$
 } \\
{c $\rightarrow$ fixed parameters}
}

Model 1: const*tbabs*(powerlaw*highecut+bbodyrad) \\
Model 2: const*tbabs*(cutoffpl+bbodyrad) \\
Model 3: const*tbabs*(NPEX+bbodyrad) \\
Model 4: const*tbabs*(powerlaw*newhcut+bbodyrad)\\
Model 5: const*tbabs*(powerlaw*fdcut+bbodyrad)\\

\end{flushleft}
\end{table*}

\begin{table*}
	\caption{Spectral fit parameters of Obs--2 with phenomenological models~(1--70~keV). }
\label{spec-para-obs2}
\begin{tabular}{ c c c c c c c c c}
\hline
\hline
Parameters & Model 1 & Model 2 &  Model 3 & Model 4  & Model 5   \\
\hline
&   &  Obs--2~(ObsID T01\_202T01\_9000001640)   &  \\
\hline

$N\rm{_{H}}$ ($10^{22}\rm{cm^{-2}}$) &  $0.72\pm0.03$ & $0.74\pm{-0.03}$ &  $0.72\pm0.03$  &  $0.72\pm0.03$     & $0.76\pm0.02$  \\ [0.1cm]
$\Gamma$  & $0.69_{-0.05}^{+0.09}$  & $0.67\pm0.08$  &  $0.39\pm0.06$  &  $0.7\pm0.1$  &    $0.95\pm0.07$              \\ [0.1cm]
$E_{cut}$ (keV)  &  $4.2\pm0.3 $ &  $17.5\pm{0.8}$  &  $12.4\pm0.2$  & $11.6_{-0.22}^{+0.21}$  &  $1~(fixed)$      \\ [0.1cm]
$E_{fold}$ (keV) &  $18\pm1 $  & -  &-  &  $18\pm1$   &   $17.9\pm0.7$                                           \\ [0.1cm]
$kT1\rm_{bb}~(keV)$  & $1.24\pm0.03$ & $1.25\pm0.02$  & $1.24\pm0.02$   & $1.28\pm0.01$   & $1.28\pm0.01$                        \\ [0.1cm]
$kT2\rm_{bb}~(keV)$  & $0.38\pm0.01$ & $0.37\pm0.01$  & $0.37\pm0.01$   & $0.38\pm0.01$   & $0.36\pm0.01$                        \\ [0.1cm]
$E_{Fe}$ (keV) & $6.9\pm0.3$ & $6.9\pm0.3$ & $6.9\pm0.3$ & $6.9\pm{0.3}$  & $6.9\pm0.3$ \\[0.1cm]
$EW_{Fe}$ (eV) & $137_{-96}^{+34}$ & $40\pm20$ & $159_{-119}^{+6}$ & $165_{-115}^{+27} $& $130_{-110}^{+15}$ \\ [0.1cm]
$N^{a}$   & $0.9^{+0.3}_{-0.2}$ & $1.1\pm0.2$ & $0.7\pm0.1$  &     $0.64\pm0.01$     &      $3.4\pm0.6$                \\ [0.1cm]
$const_{LAXPC}$  & $1.20\pm0.02$   & $1.20\pm0.02$   &  $1.20\pm0.02$  & $1.20\pm0.02$     & $1.20\pm0.02$                       \\ [0.1cm]
${Unabsorbed Flux (1-70 keV)}^b$  & $10.78\pm0.04$ & $10.78\pm0.04$ & $10.78\pm0.04$ & $10.78\pm0.04$ & $10.78\pm0.04$  \\ [0.1cm]
Reduced ${\chi}^2$~(dof)   & 0.98~(626)  &  1.00~(630)  & 0.99~(630)  & 0.99~(626)  &        1.02~(630)                   \\ [0.1cm]
\hline

\end{tabular}
\\
\begin{flushleft}
{{\bf{Note}}:
{a $\rightarrow$ Normalization~($N_{}$)
 is in units of $\rm{photons~cm}^{-2}~\rm{s}^{-1}~\rm{keV}^{-1}$ at 1~keV.} \\
{b $\rightarrow$ Unabsorbed flux in units $10^{-8} ergs \, cm ^{-2} s^{-1}$
 } \\
{c $\rightarrow$ fixed parameters}
}

Model 1: const*tbabs*(powerlaw*highecut+bbodyrad+bbodyrad+gaussian) \\
Model 2: const*tbabs*(cutoffpl+bbodyrad+bbodyrad+gaussian) \\
Model 3: const*tbabs*(NPEX+bbodyrad+bbodyrad+gaussian) \\
Model 4: const*tbabs*(powerlaw*newhcut+bbodyrad+bbodyrad+gaussian)\\
Model 5: const*tbabs*(powerlaw*fdcut+bbodyrad+bbodyrad+gaussian)\\

\end{flushleft}
\end{table*}

\section{Summary}

\begin{itemize}

\item \emph{AstroSat} observations of J0243 performed during its 2017-2018 outburst have allowed us to detect pulsations up to
150~keV.~These observations were made during two different levels of accretion viz.,~sub-Eddington~($L_{X}~{\sim}~7{\times}10^{37}~\rm{erg~s^{-1}}$) and super-Eddington~($L_{X}~{\sim}~6{\times}10^{38}~\rm{erg~s^{-1}}$).

\item Pulse profiles show a strong energy and luminosity dependence which is consistent with results from \emph{NICER} and \emph{Fermi}–\textsc{GBM}. 

\item Our study of broad-band X-ray spectra does not show any dip-like feature indicative of a cyclotron line.		
		
\item 
Spectral data from observations made at the sub-Eddington level could be modeled well using an absorbed high energy cut-off power law and a blackbody. Data obtained during the super-Eddington phase of the source, however, requires additional components such as another blackbody and a Gaussian component for the iron emission line.

\item 		
The presence of two blackbodies: one with a radius of $18-19~\rm{km}$ for the high temperature one, and another with a radius of $121-142~\rm{km}$ for the low temperature one, possibly indicates contribution to thermal emission from the accretion column and optically-thick outflows.
		
\end{itemize}

 \section*{Acknowledgments}

The authors gratefully acknowledge the referee for
his/her useful suggestions that helped us to improve
the presentation of the paper.
A.B is grateful to both the Royal Society, U.K and to
SERB (Science and Engineering Research Board), India.
A.B is supported  by  an  INSPIRE  Faculty  grant
(DST/INSPIRE/04/2018/001265)  by  the  Department  of
Science and Technology, Govt. of India and also acknowledges the financial support of ISRO under \emph{AstroSat} archival Data utilization program (No.DS-2B-13013(2)/4/2019-Sec. 2).~She is also thankful to Dr Nirmal Iyer
for offering his kind help in creating Figure~5 of this paper and to S. Bala for useful discussions. \\
For the use of \emph{AstroSat} data, we acknowledge support from ISRO for mission operations and data dissemination through the ISSDC.
We thank LAXPC POC at TIFR and 
CZTI POC at IUCAA for verfiying and releasing the data.
We are also thankful to the AstroSat Science
Support Cell hosted by IUCAA and TIFR 
 for providing the necessary data analysis software.~
 This work has used data from \textsc{SXT} which was developed at TIFR, Mumbai.
 We thank the SXT POC for verifying and releasing the data through the ISSDC data archive and for providing the necessary software tools. 
We are also very grateful to Dr Colleen A. Wilson-Hodge
for providing the pulse profiles, obtained using the \emph{NICER} and 
\emph{Fermi}--\textsc{GBM} data. 
D.A acknowledges support from the Royal Society, United Kingdom. 
The authors would also like to
thank a UGC-UKIERI Thematic Partnership for support.	\\

\section*{DATA AVAILABILITY}
The data underlying this article are publicly available in ISSDC,
at 
https://astrobrowse.issdc.gov.in/astroarchive/archive/Home.jsp

\newpage

\bibliography{complete-manuscript}{}

 \bibliographystyle{mnras}
\appendix

\section
{\textsc{LAXPC20} source+background and background spectra during Obs--1 and Obs--2}

\begin{figure*}
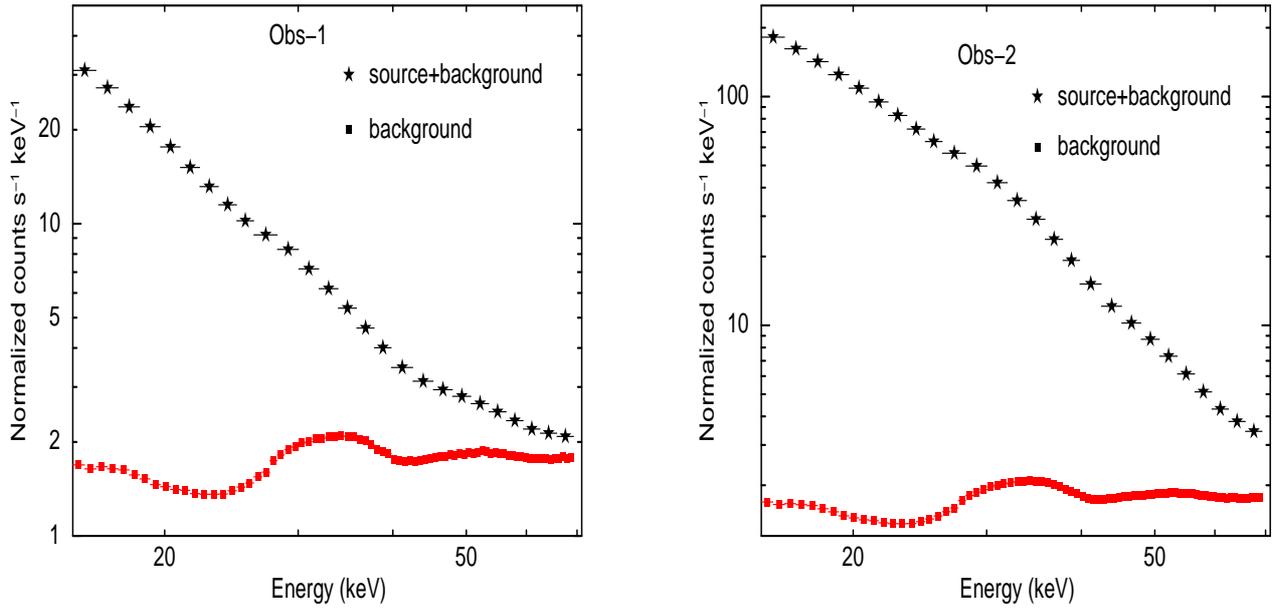

\centering
\begin{minipage}{0.45\textwidth}
\includegraphics[height=\columnwidth,width=\columnwidth, angle=-90]{fig14.eps}
\end{minipage}
\hspace{0.05\linewidth}
\begin{minipage}{0.45\textwidth}
 \includegraphics[height=\columnwidth,width=\columnwidth,angle=-90]{fig15.eps}
\end{minipage}
\caption{Plot showing \textsc{LAXPC20} source+background and background spectra simultaneously in the 15--70~\rm{keV}
	energy band. An instrumental artifact, a bump at around $\sim$33$\rm{keV}$ \citep{Antia17} can be seen in both figures with background dominance at higher energies during Obs--1 in comparison to Obs--2.~Although the signal to noise ratio indicates that the energy range can be extended to higher energies, but systematic features dominate the spectrum of Obs--1. }
\label{bkg}
\end{figure*}

\end{document}